\documentclass[10pt,aps,showpacs,preprintnumbers,twocolumn,amsmath,amssymb,prb,floatfix,superscriptaddress]{revtex4-1}

\usepackage{graphicx}
\usepackage{color}
\usepackage{subfig}

\usepackage{ifpdf}

\ifpdf
\usepackage[pdftex]{hyperref}
\else
\usepackage[hypertex,ps2pdf,dvips,colorlinks,urlcolor=blue,citecolor=blue]{hyperref}
\fi

\newcommand\be{\begin{equation}}
\newcommand\ee{\end{equation}}
\newcommand\erf[1]        {\eqref{#1}}
\newcommand{\ud}          {\mathrm d}

\newcommand\mc            {\mathcal}
\newcommand\p             {\partial}
\newcommand{\vev}[1]{\left\langle #1 \right\rangle}

\begin{document}

\title{Quantum quenches in the sine--Gordon model: a semiclassical approach}
\author{M. Kormos}
\affiliation{MTA-BME ``Momentum'' Statistical Field Theory Research Group\\
Department of Theoretical Physics, Budapest University of Technology and Economics, 1111 Budapest, Budafoki \'ut 8, Hungary}
\author{G. Zar\'and}
\affiliation{MTA-BME ``Momentum'' Exotic Quantum Phases Research Group\\
Department of Theoretical Physics, Budapest University of Technology and Economics, 1111 Budapest, Budafoki \'ut 8, Hungary}

\date{\today}

\begin{abstract}

We compute the time evolution of correlation functions after quantum quenches in the sine--Gordon model within the semiclassical approximation which is expected to yield accurate results for small quenches. We demonstrate this by reproducing results of a recent form factor calculation of the relaxation of expectation values. Extending these results, we find that the expectation values of most vertex operators do not decay to zero. We give analytic expressions for the relaxation of dynamic correlation functions, and we show that they have diffusive behavior for large timelike separation.

\end{abstract}

\maketitle

\section{Introduction}
Questions of the relaxation and thermalization of isolated quantum systems have attracted a lot of attention over the last decade \cite{Polkovnikov2011,Gogolin2015}. Under which conditions a given system relaxes or thermalizes? If the asymptotic stationary state is not thermal, can it be described within the framework of statistical physics? How quickly is the asymptotic state reached and what are the characteristics of the time evolution? The increased interest in these fundamental questions is to a major part due to the spectacular advances in cold atom experiments which are able to study the coherent evolution of isolated quantum systems, moreover, many properties and parameters of the systems are tunable \cite{Kinoshita2006,Hofferberth2007,Trotzky2011,Gring2012}.

With the scope of understanding thermalization or the lack thereof, many studies focused on the asymptotic steady state after a quantum quench \cite{CalabreseCardy2006}. Still, there are many unsettled questions. An example is the Generalized Gibbs Ensemble \cite{Rigol2007} that was proposed to describe the stationary state of integrable systems which possess an excessive number of conserved quantities. Lately, its applicability for continuum systems was questioned \cite{Kormos2013}, and its natural implementation using local conserved quantities was even shown to fail to capture the steady state of the XXZ spin chain \cite{noGGE2014,Wouters2014}.

Even less is known about the details of the relaxation process and the time scales of the relaxation. Numerical approaches 
are usually constrained either by the system size or the times until they are able to follow reliably the evolution of the system, and the long time behavior in the thermodynamic limit is very difficult to study. This is especially true for continuum many-body systems. A promising direction is based on the quench action method \cite{CauxTimeEvolution2013,DeNardis2014a,DeNardis2015} (see below). Progress in the analytic description has been made in conformal field theories \cite{CalabreseCardy2006} and in systems that can be mapped to free bosons or free fermions \cite{Sengupta2003,Cazalilla1997,Silva2008,Rossini2010,FiorettoMussardo2010,Iucci2010,Calabrese2011,Igloi2011,Divakaran2011, Rieger2011,Blass2012,Essler2012,Heyl2013,Bucciantini2014,Kormos2014, DeNardis2014a, spyrosGGE,Fagottipre2014}. 
A notable exception is Ref. \onlinecite{Pozsgay2014}.

The relaxation of quantum many body systems can happen in a number of steps. For example, a weakly nonintegrable integrable system can reach a prethermalization plateau \cite{Berges2004,Moeckel-Kehrein2008,Kollar2011,Gring2012,Essler-Kehrein2013,Fagottipre2014} close to the steady state of the integrable model and real thermalization takes place only at much larger timescales.

Obtaining analytical results for genuinely interacting systems is notoriously hard even in integrable models, 
where the spectrum and the matrix elements of local operators, the so-called form factors, are usually known.
Based on these ingredients, 
a linked cluster expansion can be constructed where the small expansion parameter is essentially the density of excitations after the quench \cite{Schuricht2012,Bertini2014}. 
Summing up the series is a daunting task, nevertheless, some progress can be made at least regarding the short time behavior after a quantum quench. 
Another possible approach is based on the so-called quench action technique for Bethe Ansatz integrable systems \cite{CauxTimeEvolution2013} which works directly in the thermodynamic limit.  
Both methods have been applied to study the relaxation dynamics in the sine--Gordon model in Ref. \onlinecite{Bertini2014}, and predicted an exponential decay of the vertex operator  $\vev{e^{i\beta\Phi(x,t)/2}} \sim e^{-t/\tau}$.

Here we shall follow a complementary and more intuitive approach, and study the evolution of dynamical correlations 
after a quantum quench by extending the \emph{semiclassical approach} of Refs.
\cite{Sachdev1996,Sachdev1997,Damle1998,Rapp2006,Rapp2008,Damle2005,Rapp2008,
Rossini2010,Igloi2011,Divakaran2011,Rieger2011,Blass2012,Evangelisti2013} to study small quenches in the gapped phase of the sine--Gordon model.
The sine--Gordon model is a paradigmatic model providing the low energy effective description of a wide range of one-dimensional systems  including spin chains, spin ladders, and cold atomic gases \cite{Giamarchibook,SachdevBook,Cazalilla_bosonizing2003,Buchler2003,Essler2005,Gritsev2007a}. 
It is defined by the action
\be
\mathcal{S} = \frac{c}{16\pi} \int\ud x\ud t\left[\frac1{c^2}(\p_t\Phi)^2-(\p_x\Phi)^2+ \lambda \cos(\beta\Phi)\right]\,,
\label{eq:action}
\ee
with $\Phi(x,t)$ a bosonic field. In this work, for the sake of simplicity,  we shall focus on the gapped  repulsive phase ($1/\sqrt{2}<\beta<1$),  where the model contains massive topological excitations (kinks), the so-called solitons and antisolitons with charge $m=\pm1$ interpolating between neighboring minima of the $\cos(\beta\Phi)$ potential, but no bound states (breathers) exist.\footnote{In contrast to Ref.\cite{Gritsev2007,Gritsev2007a}, the presence of breathers is not expected to influence the results presented here.}
 
The time evolution of the vertex operator $\vev{e^{i\beta\Phi}}$ was studied after quenches in the attractive regime in Ref. \onlinecite{Gritsev2007}. Analytic results have been derived for the correlations of the same operator for quenches between the exactly solvable points $\lambda=0$ and $\beta=1/\sqrt{2}$ in Ref. \onlinecite{Iucci2010}. In Ref. \onlinecite{bla10} the time evolution of an inhomogeneous initial state was studied.

The  semiclassical  method  is based on the observation that at small temperatures or after a small quench the density of quasiparticles as well as their velocity is small. Therefore quasiparticles can be treated as classical entities apart 
from collisions, where their de Broglie wavelength becomes comparable to their separation. These quantum effects 
are taken into account by using the low energy limit of the two-particle scattering matrix. 

The semiclassical approach turned out to yield a remarkably accurate description of many gapped systems. It has been successfully applied to compute finite temperature correlation functions \cite{Sachdev1996,Sachdev1997,Damle1998,Damle2005,Rapp2006,Rapp2008}, as well as the time evolution of correlations after global and local quenches \cite{Rossini2010,Rieger2011,Divakaran2011,Blass2012,Evangelisti2013}.

As we  demonstrate here, this approach is able to capture the leading behavior of the decay processes and reproduces the tediously obtained results of Ref.~\onlinecite{Bertini2014}  with ease.

However, we can go significantly beyond these results. While extending the form factor based calculations of Ref. \onlinecite{Bertini2014} to other observables (e.g. two-point functions) seems to be a very demanding task, in the semiclassical method it poses only slight, surmountable complications. Assuming that only soliton-antisoliton pairs are present in the initial state (as is the case for fixed $\Phi$ initial conditions), we obtain new results in two directions. On the one hand, we calculate the relaxation of general vertex operators 
\be
\label{Gdef}
G_\alpha(t)=\vev{e^{i\alpha\Phi(x,t)}} = \langle\psi_0|e^{iHt}e^{i\alpha\Phi(x,0)}e^{-iHt}|\psi_0\rangle
\ee
with the somewhat surprising result that --- in the universal limit --- they do not decay to zero but approach finite asymptotic values. \footnote{The reason and the limitations of this surprising result shall be discussed in the concluding section of this work.} 
On the other hand, we compute the time evolution of dynamical two-point functions of general vertex operators
\be
\label{Cdef}
C_\alpha(x'-x;t,t')=\vev{ e^{i\alpha\Phi(x,t)}e^{-i\alpha\Phi(x',t')} }\,.
\ee
We show that the two-time correlations show diffusive behavior for generic values of $\alpha.$ 
This is expected to some extent given that diffusive behavior was also observed in 
the semiclassical treatment of correlations in thermal equilibrium \cite{Damle2005,Rapp2006,Rapp2008}.

The paper is organized as follows. In Section II we describe the semiclassical approach in detail. Expectation values of vertex operators are computed in Section III. We calculate the time evolution of general dynamic correlation functions  
in Section IV, and we analyze the equal time correlations, the local correlation functions and the correlations in the asymptotic state separately. We give our conclusions in Section V.

\section{The semiclassical method}

\begin{figure}
{\includegraphics[width=.3\textwidth]{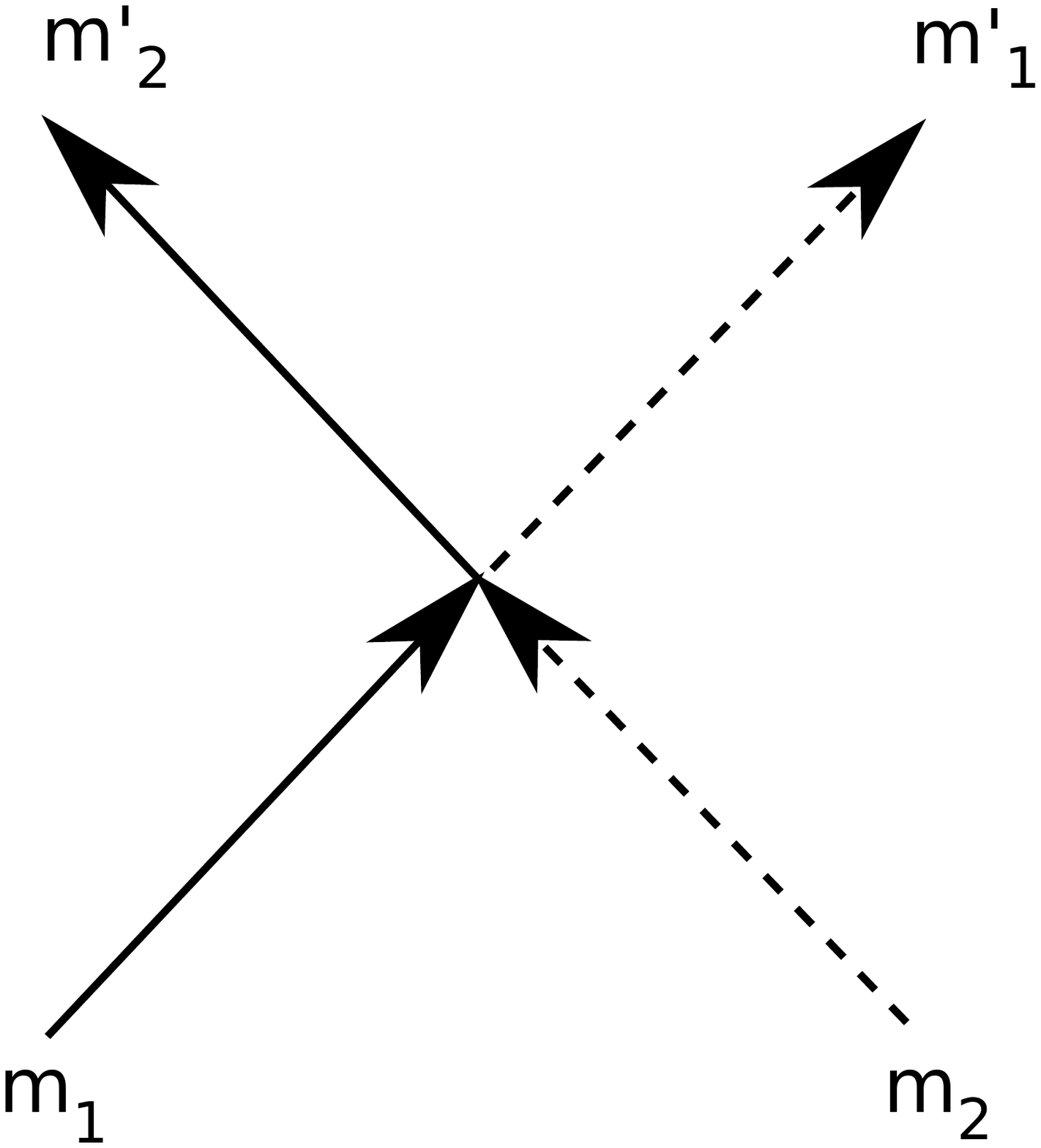}}\
\caption{A two-body collision of kinks.}
\label{fig:Sm}
\end{figure}

\begin{figure}[!t]
 {\includegraphics[width=.45\textwidth]{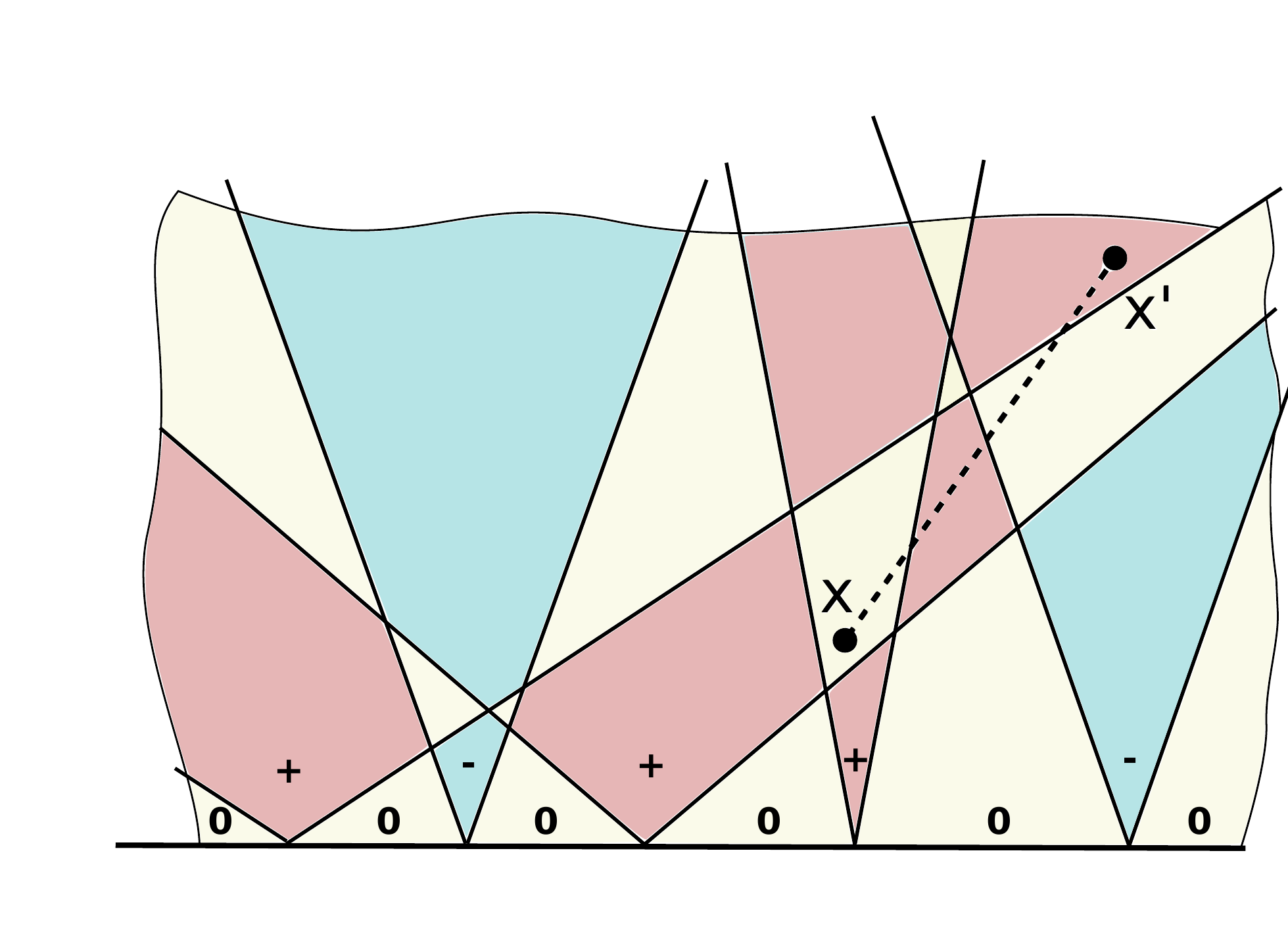}}
  \caption{A kink configuration for initial states with soliton-antisoliton pairs only. In this example $n_L=1,n_R=1,n_I=1,n_A=1.$}
 \label{fig:domains}
\end{figure}

As discussed in the Introduction,  for small quenches, the energy density injected in the system is small, and quasiparticles are generated with a low density and with energies  close to the energy gap. In this low density limit, the motion 
of quasiparicles is `slow' and can be treated semiclassically.
The quantum expectation values \erf{Gdef} and \erf{Cdef} are calculated as averages over the kink configurations, that is over the initial positions, velocities and charges of the kinks \cite{Sachdev1996}. 
Importantly, due to momentum and energy conservation,  the trajectories of quasiparticles remain straight lines in 1D and 
follow ``rays'',   (see  Figs. \ref{fig:Sm},\ref{fig:domains}). 

In the particular case of the repulsive sine-Gordon model, 
in the semiclassical limit, most of the time the field $\Phi$ remains close to 
 minima of the cosine term in  Eq.~\eqref{eq:action} 
\be
\Phi = n\frac{2\pi}\beta\,,\qquad\qquad n\in \mathbb{Z}\,.
\ee
Quasiparticles are just kinks (domain walls) separating 
domains of constant $\Phi$, such that $n$ increases (decreases) by one when crossing 
in the positive spatial direction a soliton (antisoliton) trajectory.

In the small density limit only  two-particle collisions are relevant.  Given the small velocity of quasiparticles, 
the scattering matrix of solitons and antisolitons can be approximated by its low momentum limit as 
\be
\label{lowS}
S_{m_1',m_2'}^{m_1,m_2} = (-1)\,\delta_{m_1,m_2'}\delta_{m_2,m_1'}\,,
\ee
i.e. kinks scatter as ``hard balls'' (c.f. Fig. \ref{fig:Sm}). 
This structure is crucial for the rest of this work: it
implies that the spatial sequence of the topological charges of the kinks (solitons and antisolitons) remain the same for all 
times in this asymptotic limit. Alternatively, in terms of domains, Eq.~\eqref{lowS} implies that the `color' \emph{sequence}
of domains remains invariant under time evolution (see Fig.~\ref{fig:domains}).

As discussed above, any quantum quench of the Hamiltonian generates a gas of quasiparticles. 
However, very importantly, in many cases the post-quench state seems to possess a particular, correlated structure in terms of quasiparticle pairs. Following Ref. \onlinecite{Bertini2014}, we take our initial state to be
\be
\label{psi0}
|\psi_0\rangle = \exp\left\{\int_0^\infty\frac{\ud\theta}{2\pi} K_{mm'}(\theta)\hat Z_m^\dag(-\theta)\hat Z_{m'}^\dag(\theta)\right\}
|0\rangle\,,
\ee
where $\hat Z^\dag_m(\theta)$ creates a kink of type $m=\pm$ with relativistic rapidity $\theta.$ 
These states are coherent superpositions of kink pairs and are called integrable initial states due to their resemblance to  
integrable boundary states~\cite{FiorettoMussardo2010}. The exponential form featuring a single creation amplitude 
(for each pair type) generates \emph{independent} pairs of particles with opposite velocities.
For small quenches,  the pair creation amplitudes $K_{mm'}(\theta)$ and the corresponding densities  are small, 
\be
\rho_{mm'} \approx \int_0^\infty\frac{\ud\theta}{2\pi}M c\cosh(\theta) |K_{mm'}(\theta)|^2\ll1\,,
\ee
where $M$ is the kink mass \footnote{The kinks mass $M$ can be expressed in terms of the parameters of the Hamiltonian $\lambda,\beta,c;$ its precise form is not necessary for our discussion.} .

For small quenches, the form \erf{psi0} of the initial state can be justified on very general grounds: by momentum conservation, a local perturbation, to the lowest order, gives rise to pairs of kinks flying away from each other with the same velocity. Moreover, if the field is originally constant (Dirichlet boundary condition) then, since the field must remain unaltered away from the perturbation, the total topological charge of each pair must be zero, i.e. pairs must form soliton-antisoliton pairs. Since a homogeneous global quench by the integral of a local operator is a sum of such local quenches, thus the post-quench state will be populated by independent pairs. Indeed, this pattern of the quench creating pairs of quasiparticles has been observed for several integrable systems and initial states~\cite{Silva2008,Sotiriadis2010a,deNardis2014,Pozsgayoverlaps2013,BrockmannGeneralizedGaudin2014B,Sotiriadis2014}, even for large quenches.
The variety of quenches featuring the pair structure suggests that this may be a general phenomenon. We note that the sine-Godon model with the initial state \erf{psi0} appears in the description of interference patterns between split 1D condensates and the Ramsey sequence of two-component 1D bosons\cite{Kitagawa2011,Foini2014}.

We thus assume that the post-quench semiclassical configuration consists of a collection of uniformly and independently distributed pairs of straight lines, placed along the $t=0$ axis, 
 and a random sequence of soliton-antisoliton charges (see  Fig. \ref{fig:domains}).  Kink pairs have a velocity distribution $f_{mm'}(v)$ ($v>0$), i.e. each pair of kinks with charges $m$ and $m'$ traveling with velocities $-v$ and $v$ is created with a probability density $f_{mm'}(v).$ For the specific initial state \erf{psi0} this distribution is related to the amplitudes $K_{mm'}(\theta)$ as 
\be
\label{fKrel}
f_{mm'}(v) \approx \frac{M}{2\pi\rho}|K_{mm'}(v/c)|^2 \,,
\ee
where $\rho=\sum_{m,m'}\rho_{mm'}$ is the total density of pairs and we used that the velocities are nonrelativistic and $v/c=\tanh\theta\approx\theta.$ With this definition, the distribution functions are normalized as
$\sum_{mm'} \int_0^\infty\ud v f_{mm'}(v) = \sum_{mm'}p_{mm'} =1$.
%

In our case, only soliton-antisoliton and antisoliton-soliton pairs are created with equal probability, thus $K_{++}=K_{--}=0$ and $K_{+-}=K_{-+}=K.$  We can therefore characterize the velocity distribution of the kinks by a single function, 
\be
f(v)\equiv \frac{M}{\pi\rho}|K(v/c)|^2,\phantom{nnn} \int_0^\infty\ud v f(v) =1.
\ee

Our goal is to determine the time evolution of the expectation value \erf{Gdef} and the correlation function \erf{Cdef}. 
Correlation functions depend on the difference $\Phi(x_1,t_1)-\Phi(x_2,t_2)$ which, apart from a prefactor 
$2\pi /\beta$ is equal to the sum of charges of the kinks that cross the straight line connecting the two 
points where the operators are inserted. 
As noticed earlier, due to the special structure of the universal S-matrix \eqref{lowS}, 
 the spatial sequence of domains of constant $\Phi$ remains unchanged under the time evolution (see Fig. \ref{fig:domains}). 
 Domains can thus be labeled by an integer, and $\Phi(x,t)=\Phi_l$ in case  the point $(x,t)$ lies in the $l^\text{th}$ domain. Consequently, $\Phi(x_1,t_1)-\Phi(x_2,t_2)=\Phi_{l_1}-\Phi_{l_2}$ if $(x_1,t_1)$ and $(x_2,t_2)$ lie in the $l_1^\text{th}$ and $l_2^\text{th}$ domain, respectively, and 
\be
\label{phidiff}
\Phi_{l_1}-\Phi_{l_2} = \frac{2\pi}\beta \sum_{i=l_1+1}^{l_2} m_i\,,
\ee
where $\{m_i\}$ are the charges of the $s=|l_2-l_1|$ domain walls (kinks) encountered while 
going from domain $l_1$ to $l_2.$ 
The precise value of $s$ is determined by the number and directions of the rays that intersect the segment between the two operator insertion points, 
and an averaging over the initial positions $\{x_i\}$ and velocities $\{v_i\}$ of the pairs as well as
over the charges $\{m_i\}$ 
needs to be performed. The correlation function \erf{Cdef} is thus expressed as
\be
C_\alpha(\Delta x;t,t')=\mc{C}^{(0)}_\alpha(\Delta x,t'-t) \left\langle e^{i\frac{2\pi}\beta \sum_{i=l_1+1}^{l_2} m_i
}\right\rangle_{\{m_i\},\{(x_i,v_i)\}}\,,
\ee
with $\mc{C}^{(0)}_\alpha$  denoting the vacuum correlation function.

\section{Relaxation of expectation values}
\label{sec:exp}

Let us  start by calculating the time evolution of the expectation value $\vev{e^{i\alpha\Phi(x,t)}}$.
For this we need to determine the indices of domains
 in which the points $(x,t)$ and $(x,0)$ lie. Since $\Phi(x,t=0)=0$, however, it suffices to know the number of domains $s$ we shift to the left or to the right while we travel along the straight vertical segment $\mc{S}$ connecting the points 
 $(x,0)$ and $(x,t).$ 
 The number $s$ is given by the difference of the numbers of rays intersecting the segment $\mc{S}=[(x,0),(x,t)]$
  from the right and from the left, $s=n_{+}-n_{-}.$ 

Clearly, at most one ray from each pair can intersect the vertical segment, and a  ray of velocity $v>0$ intersecting the segment from the left must be the right member of a pair originating from the spatial interval $[(x-vt,0),(x,0)].$ 
Since pairs are created uniformly at $t=0,$ the probability that the right going ray of a given pair with velocity $v$ intersects the segment from the left is $vt/L,$ where $L$ is the size of the system. The probability that a given pair leads to such an intersection is
\be
p = \int_0^{\infty}\ud v \frac{vt}Lf(v)\,.
\ee
By symmetry, the probability of left intersections is the same.  Since pairs are created uniformly, 
left and right intersections are independent Poissonian processes, and the probability of a pair configuration with $n_+$ ($n_-$) crossings from the right (left) is simply given by
\be
p(n_+,n_-) = \frac1{n_+!}\frac1{n_-!}Q^{n_++n_-}e^{-2Q}
\ee
with
$ Q = t\rho\int_0^{\infty}\ud v\, v f(v) = \vev{n_\pm}$, and  $\rho=N/L$  the total density of pairs.

Since at time $t=0$ the soliton-antisoliton pairs shrink to single points, the domain $l_1$ of $\Phi(x,0)$ lies between two pairs with probability 1 and $l_1$ is even. If $s=n_+-n_-$ is even, then so is the domain $l_2=(l_1+s)$.  In this case  domains $l_1$ and $l_2$ are separated by $s/2$ soliton-antisoliton pairs and  have necessarily the same $\Phi$ values: $\Phi_{l_2}-\Phi_{l_1}=\Phi(x,t)-\Phi(x,0)=0.$ If $s$ is odd, however, then the $l_2^\text{th}$ domain lies at $t=0^+$ in a domain between the members of a pair in which $\Phi(x,t)=\Phi(x,0)\pm 2\pi/\beta$ with equal probability. Averaging over the two possibilities gives $(e^{i2\pi\alpha/\beta}+e^{-i2\pi\alpha/\beta})/2=\cos(2\pi\alpha/\beta).$ The final result for the expectation value is then
\begin{widetext}
\be
\label{1ptres}
\frac{\vev{e^{i\alpha\Phi(x,t)}}}{\mc{G}_\alpha} = \sum_{n_+,n_-=0}^\infty p(n_+,n_-) \left(\frac{1+(-1)^{n_+-n_-}}2+\frac{1-(-1)^{n_+-n_-}}2\cos(2\pi\alpha/\beta)\right)\,,
\ee
\end{widetext}
where $\mc{G}_\alpha=\vev{e^{i\alpha\Phi(x,t)}}_\text{vac}$ is the vacuum expectation value computed exactly in Ref. \onlinecite{Lukyanov1997}.  
Carrying out the summation yields then  
\be 
{\vev{e^{i\alpha\Phi(x,t)}}}/{\mc{G}_\alpha} 
=\cos^2(\pi\alpha/\beta)+\sin^2(\pi\alpha/\beta)e^{-t/\tau}\,,
\ee
with the characteristic time $\tau$ expressed as
\be
\label{taudef}
\tau^{-1} = 4\rho\int_0^\infty\ud v v f(v)\,.
\ee
The expectation value thus  approaches an $\alpha$-dependent constant exponentially fast, 
with the relaxation  time  independent of $\alpha$, that is, independent of the operator measured. 
As a matter of fact, this time scale is, up to a $\mc{O}(1)$ constant, given by the mean distance between the particles $\rho^{-1}$ divided by the average velocity, that is, the average time between two collisions.

The somewhat surprising  non-zero asymptotic value can be understood as follows. As explained earlier, domains conserve their `colors' 
even upon collisions. Therefore, at any time, half of the domains must 
have phase $\Phi=0$, while one quarter of them possess phases  $\Phi=\pm 2\pi/\beta$, respectively.
For times $t\gg\tau,$ solitons collide randomly and each individual  domain  exhibits a random Brownian 
motion. Therefore, at any given spatial point we find with probability 
1/2 a phase $\Phi=0$ while phases $\Phi=\pm 2\pi/\beta$ occur with probabilities 1/4 each. This immediately 
yields the asymptotic expectation value  $\vev{e^{i\alpha\Phi(x,t)}}/{\mc{G}_\alpha}\to \cos^2(\pi\alpha/\beta)$.

Remarkably, expanding the result \erf{1ptres} for small $t$ we recover the results of 
Ref. \onlinecite{Bertini2014}, 
\be
\frac{\vev{e^{i\alpha\Phi(x,t)}}}{\mc{G}_\alpha} = 1 - \sin^2(\pi\alpha/\beta)\,\frac{t}\tau+\mc{O}(t^2)\,,
\ee
 obtained through the linked cluster expansion.  There, however,
 higher order contributions being missing, a pure exponential decay to zero was assumed and an operator dependent decay rate, $\tau_\alpha^{-1}=\sin^2(\pi\alpha/\beta)\tau^{-1},$ was defined.

In the special case  $\alpha=\beta/2+k\pi,$ $k\in \mathbb{Z},$ the asymptotic value is zero and a pure exponential decay is obtained.
For these operators we can compute the expectation value even without assuming $f_{++}(v)=f_{--}(v)=0,$ i.e. in the presence of soliton-soliton and antisoliton-antisoliton pairs. In this case  kinks crossing the segment $\mc{S}$ simply flip the sign of 
$e^{i\alpha\Phi(x,t)}$ yielding 
\be{\vev{e^{i\beta\Phi(x,t)/2}}} = {\mc{G}_{\beta/2}} \:\,\, e^{-t/\tau}\,,
\ee
where now $f(v)=\sum_{mm'}f_{mm'}(v).$ We thus recover the exponential decay found in Ref. \onlinecite{Bertini2014}
(c.f. Eq.~\erf{fKrel}). Notice that in Ref. \onlinecite{Bertini2014} the representative state approach was based on a steady state computed in the leading order in the kink density but the series describing the time evolution was resummed. It is thus a non-trivial 
fact that the semiclassical approach completely reproduces the result.

For $\alpha=\beta+k\pi$ we get $\vev{e^{i\beta\Phi(x,t)}}={\mc{G}_{\beta}},$ so the operator $e^{i\beta\Phi}$ does not evolve in time in the semiclassical approximation. This is probably due to the fact that it is the interaction energy density and the kinetic energy given by the derivatives are suppressed in the small quench limit. We note that the form factor calculation  \cite{Bertini2014} leads to the same trivial result.


\section{Relaxation of correlation functions: derivation}

\subsection{Case $\alpha=\beta/2$}
\label{sec:corrbeta2}

Before turning to the general case, let us compute the dynamical two-point function 
\eqref{Cdef}
of the operator $e^{i\beta\Phi/2}$ with no restriction on the type of pairs. 
As we have seen in the previous section, this is a particularly simple case since
\be
e^{i\beta/2[\Phi(x,t)-\Phi(x',t')]} = (-1)^{\sum_{i=1}^s m_i}=(-1)^n\,,
\label{eq:which}
\ee
where $n$ stands for  the total number of trajectories intersecting the segment $\mc{S}=[(x,t),(x',t')]$ connecting the two operator insertion points. This feature is the reason why we can allow general initial states with all possible kinds of kink pairs.

Pairs of which both rays intersect the segment do not contribute.
Let us compute the probability $q$ that exactly one ray of a given pair crosses the segment, as shown in Figs \ref{fig:L}, \ref{fig:R} and \ref{fig:Ltilde}. Without  loss of generality, we shall assume that $x'>x.$ It will be useful to define the velocities
\be
\tilde v = \frac{x'-x}{t'+t}\,,\qquad v_s = \frac{x'-x}{t'-t}\,.
\ee
It is  simple  to check that if the velocity of the pair satisfies $v<\tilde v$ then both rays of a pair can cross $\mc{S}$. Then the real space domain where pairs with just one crossing ray can originate consists of two intervals, $[x-vt,x+vt]$ and 
 $[x-vt',x+vt']$ of total length $2v(t'+t)$. In the opposite case,  $v>\tilde v,$ at most one ray of a pair can cross $S.$ One of the two intervals in this case is $[x-vt,x'-vt]$ for $v<v_s$ (Fig. \ref{fig:L}) or $[x-vt,x'-vt]$ for $v>v_s$ (Fig. \ref{fig:Ltilde}), 
and the other one is $[x+vt,x'+vt']$ (Fig. \ref{fig:R}). The lengths of the intervals are $|(x'-x)-v(t'-t)|$ and $|(x'-x)+v(t'-t)|,$ where the modulus  ensures that the expressions cover both the $t'>t$ and $t'<t$ cases. Thus 
the probability that one of the two rays of a pair with velocity $v$ intersects the segment is
\be
q_v = \Theta(\tilde v-v) \frac{2v(t+t')}L +\Theta(v-\tilde v) \frac{|\Delta x-v\Delta t|+|\Delta x+v\Delta t|}L\,,
\ee
with $\Theta(x)$  the Heaviside function, $\Delta x=x'-x,$ $\Delta t=t'-t.$ Then the probability that only one ray of a given pair will cross is $q=\int_0^\infty \ud v f(v)q_v$, and the weight of a configuration having $n$ crossings has Poisson statistics, 
$p(n) = \frac1{n!} Q^n e^{-Q}$, 
with $Q$ expressed as 
\begin{multline}
\label{Qdef}
Q(\Delta x;t,t') = Nq = 2\rho(t+t')\int_0^{\tilde v} \ud v f(v)v \\
+ 2\rho\Delta x\int_{\tilde v}^{v_s} \ud v f(v) + 2\rho|t'-t|\int_{v_s}^\infty \ud v f(v) v 
\end{multline}
and $f(v)=\sum_{ab}f_{ab}(v)$. 
The correlator is then obtained by averaging  \eqref{eq:which} over all values of $n$, yielding 
\be
\label{beta2res}
C_{\beta/2}(\Delta x;t,t') =  
\mathcal{C}^{(0)}_{\beta/2}(x'-x;t'-t)\; 
e^{-2Q(x'-x,t,t')}\,,
\ee
with $\mathcal{C}^{(0)}_{\beta/2}(x'-x;t'-t)$
denoting  the vacuum correlator. 

As we noted before, this expression should also describe the correlation of the order parameter in the continuum Ising field theory. Indeed, it agrees with the scaling limit of the exact result \cite{Essler2012} for the transverse field Ising chain once an appropriate $f(v)$ function is used in it. Naturally, it can also be obtained by taking the scaling limit of the semiclassical result for the Ising chain presented in Ref. \onlinecite{Evangelisti2013}. It is important to note that it is not {\it a priori} obvious that the scaling limit of the post-quench behavior of the spin chain coincides with the post-quench behavior of the field theory, because a sudden quench can excite high energy states. In the semiclassical method it is however almost automatic, as the derivations are essentially identical, and only the dispersion relation and the distribution function differ. So, at least for quenches in the semiclassical regime, it appears that the field theory captures correctly the non-equilibrium behavior of the spin chain. This was also observed in Ref. \onlinecite{Schuricht2012}, where it was found that the asymptotic time evolution of the order parameter after a mass quench within the paramagnetic phase of the Ising field theory agrees with the scaling limit of the time dependent magnetization of the Ising spin chain after quenching the magnetic field.

\subsection{General $\alpha$}
\label{sec:general}

Let us turn to the calculation of the time dependent correlation function \eqref{Cdef} for general $\alpha$.
For the sake of simplicity, let us assume that $x'\ge x$,  $t'\ge t$, since the other cases follow by symmetry. 
Unlike the previous subsection, we now restrict the analysis to the case when there are only soliton-antisoliton and antisoliton-soliton pairs present with the same velocity distribution. As we mentioned, initial states with fixed $\Phi$ values fall into this class.

\begin{figure}[!t]
  \centering
  \subfloat[][Avoiding pair (LL).]{\includegraphics[width=.2\textwidth]{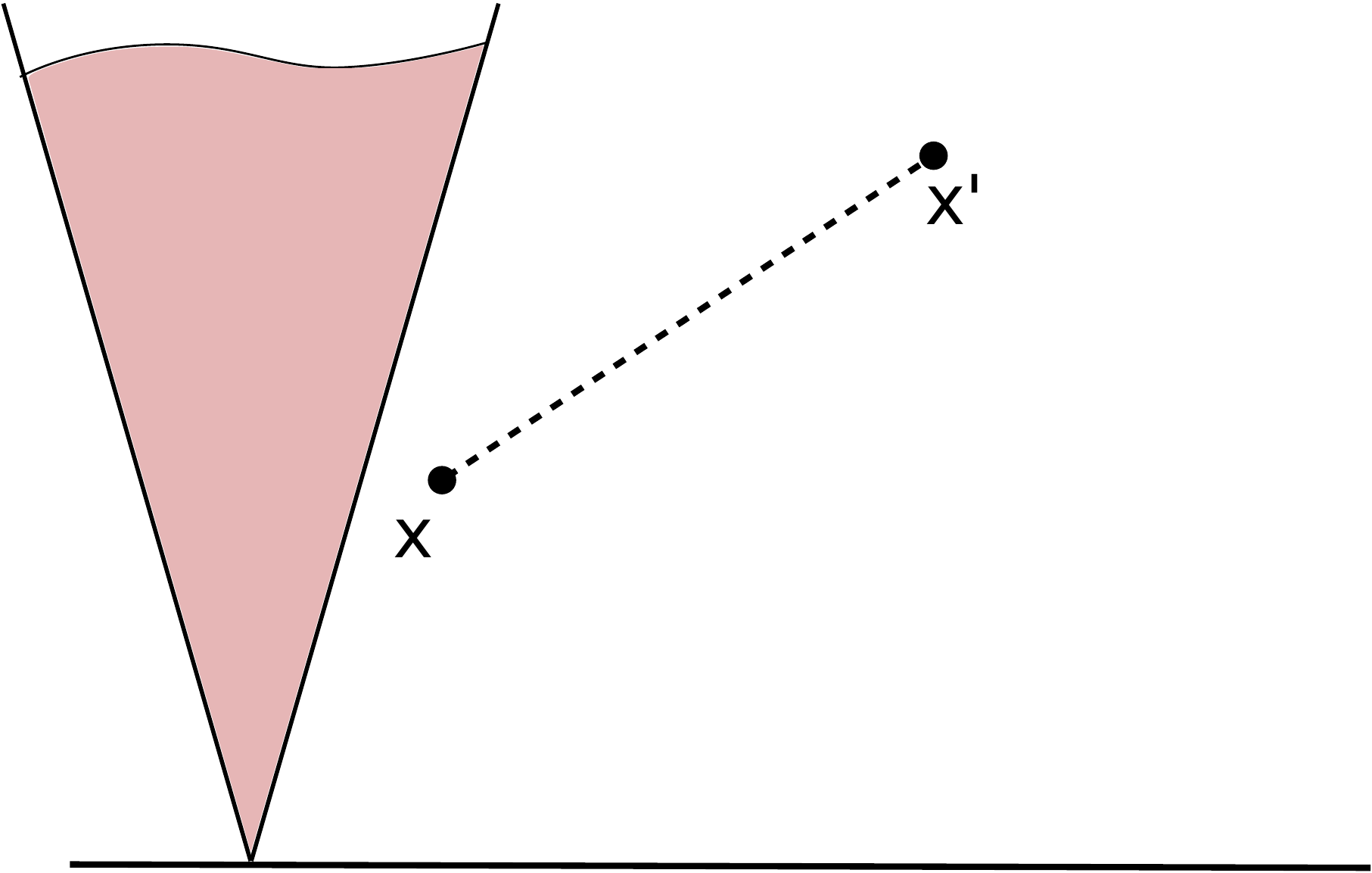}\label{fig:LL}}\qquad
  \subfloat[][Left cut (L).]{\includegraphics[width=.2\textwidth]{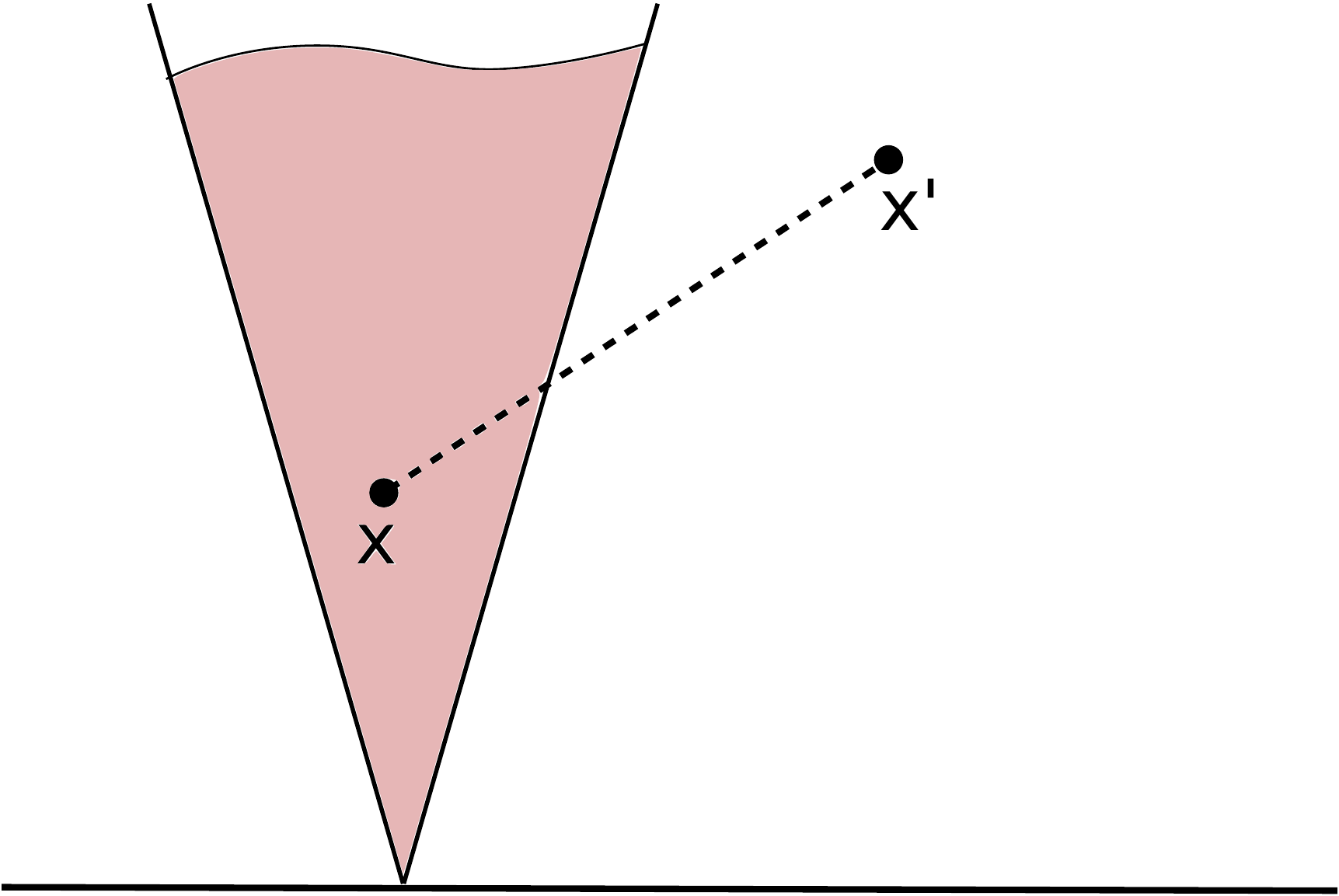}\label{fig:L}}\\
  \subfloat[][Right cut (R).]{\includegraphics[width=.2\textwidth]{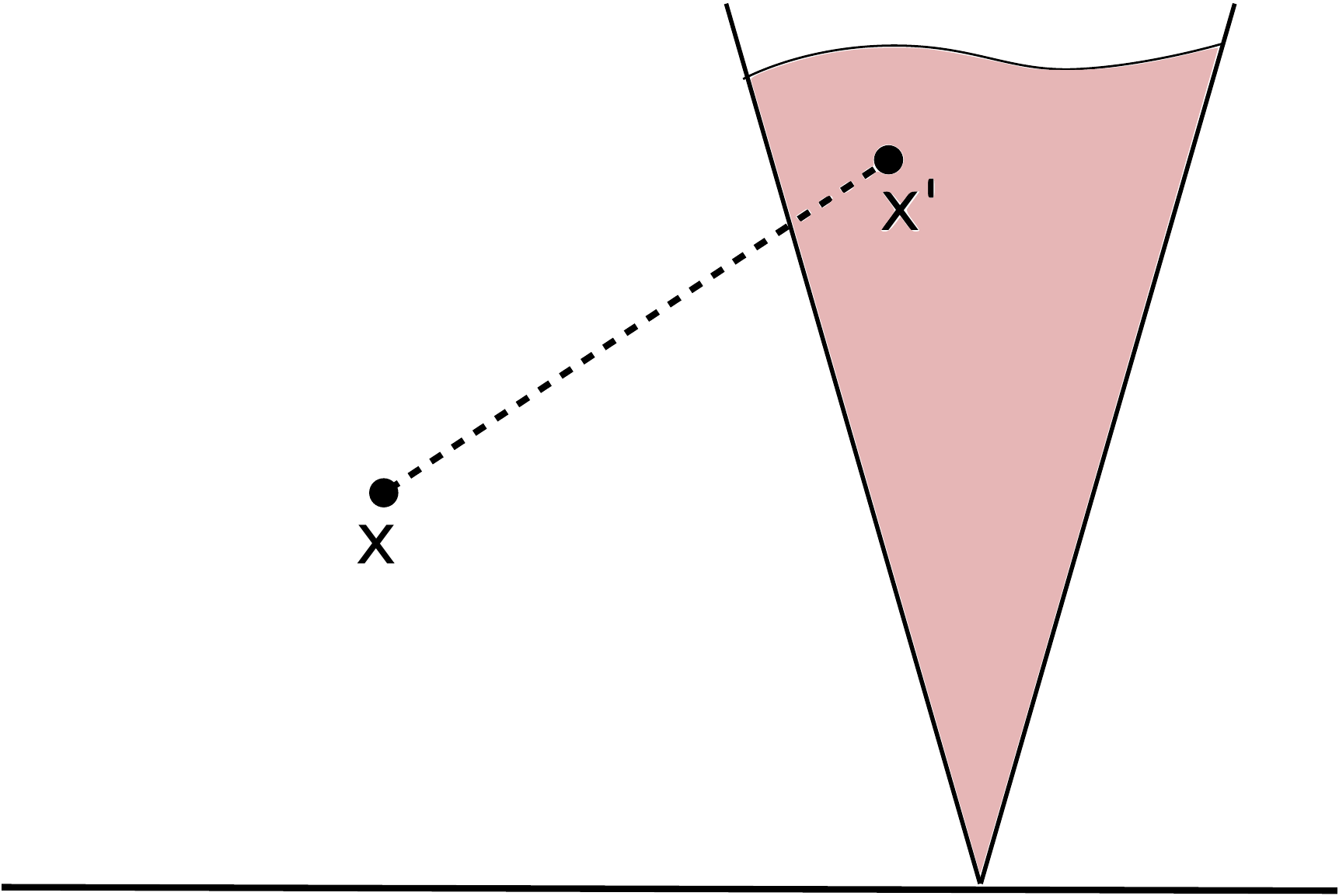}\label{fig:R}}\qquad
  \subfloat[][Double cut (D)]{\includegraphics[width=.2\textwidth]{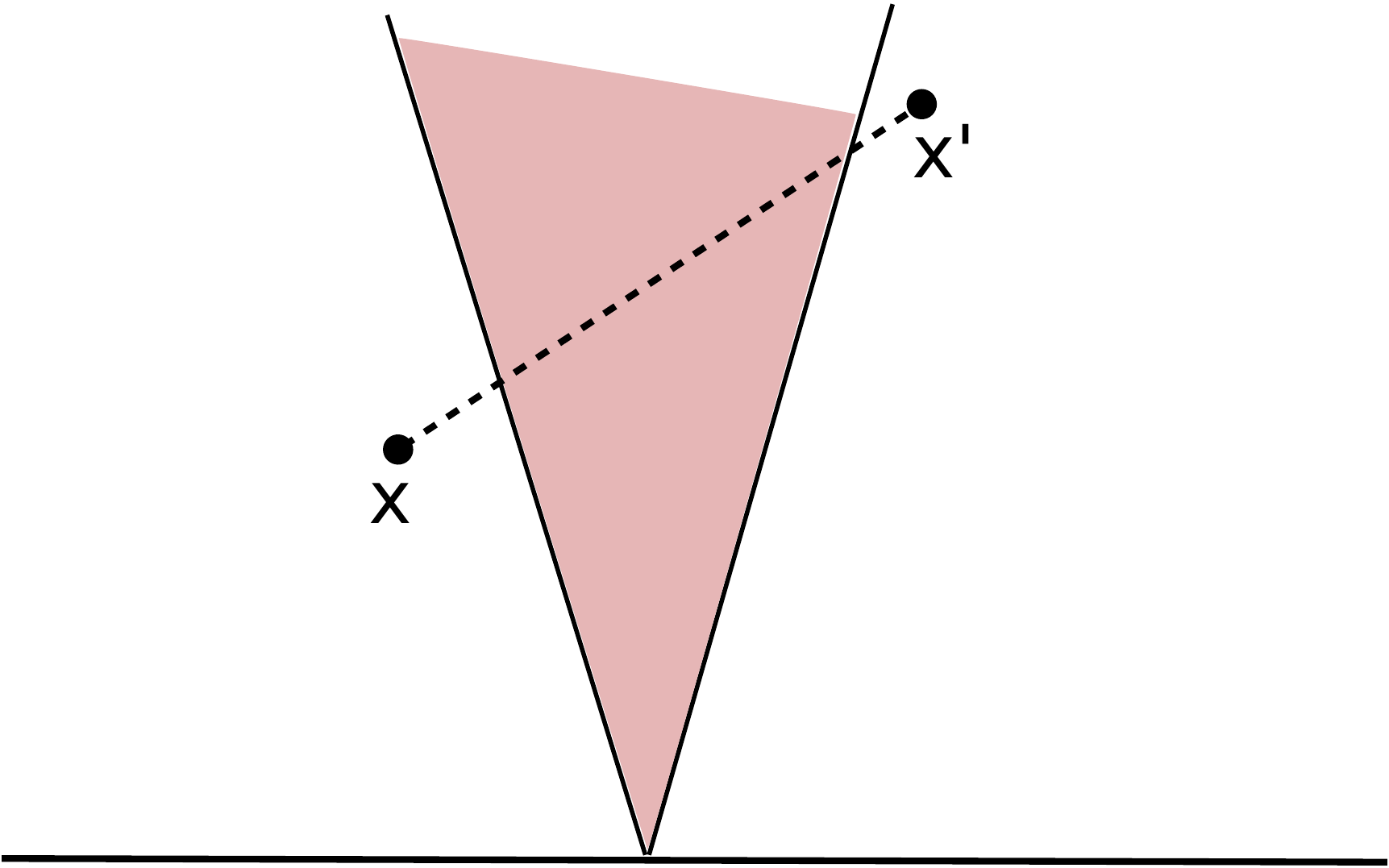}\label{fig:D}}\\
  \subfloat[][Inclusion (I).]{\includegraphics[width=.2\textwidth]{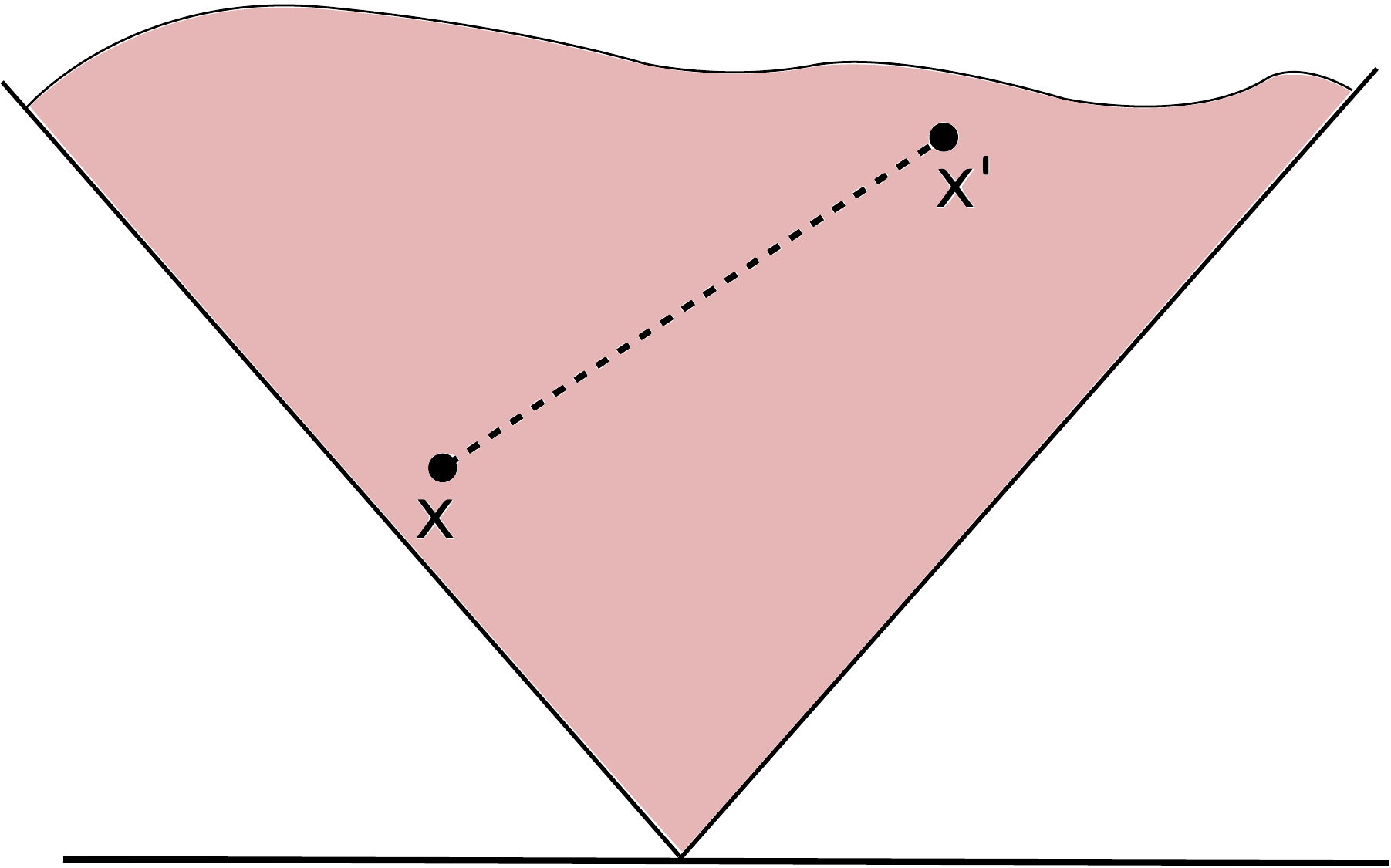}\label{fig:I}}\qquad
  \subfloat[][Left cut ``from above'' (A).]{\includegraphics[width=.2\textwidth]{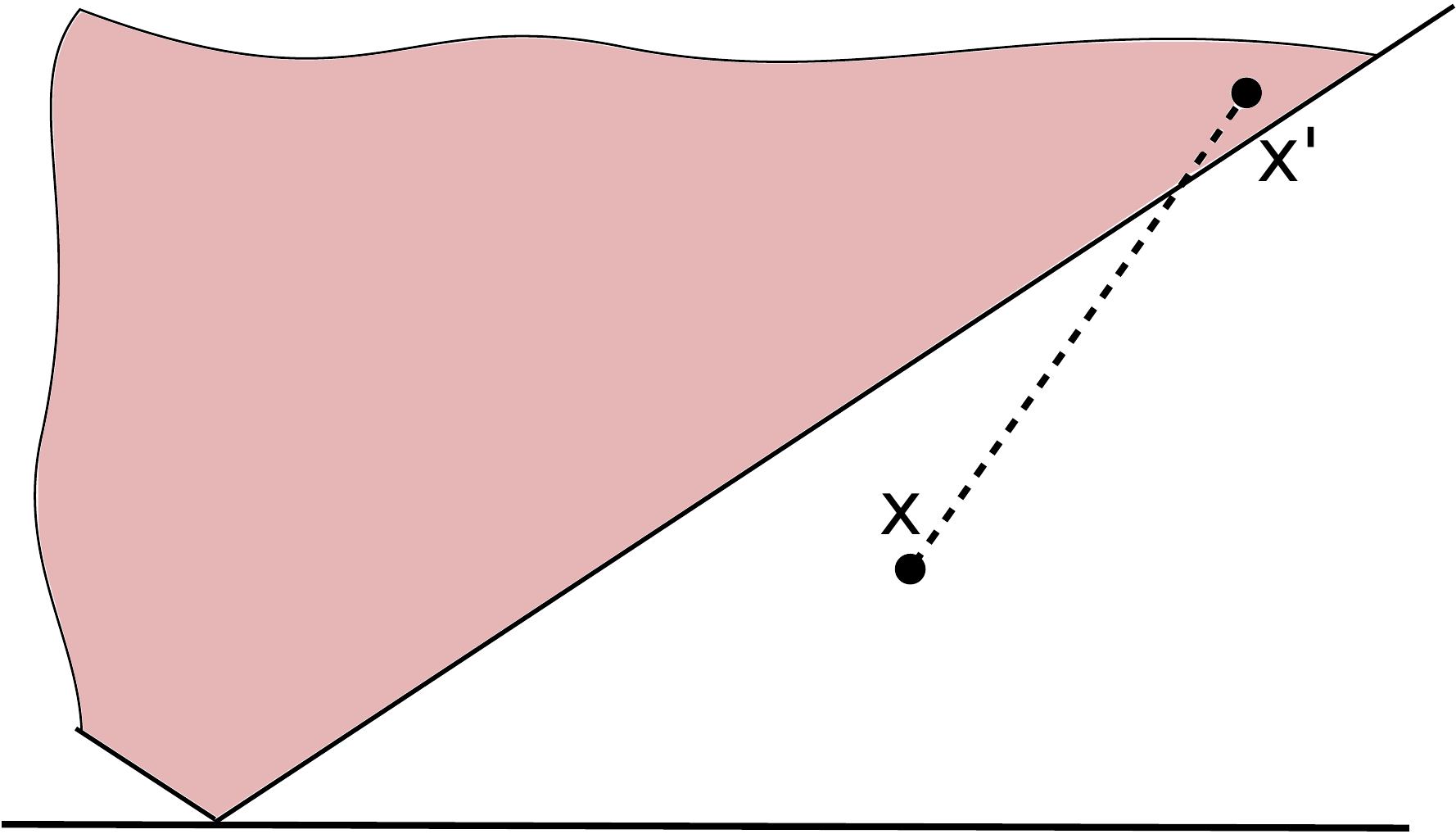}\label{fig:Ltilde}}
  \caption{Representative examples of the classes of pairs defined in the main text.}
  \label{fig:LRDI}
\end{figure}

For a generic vertex operator ($\alpha$ arbitrary), the calculation of the correlation function is similar to that of the 
expectation value \eqref{Gdef}, but technically considerably more involved. 
According to Eq. \erf{phidiff} we need to determine the distribution of the number and orientation of domains
between point $(x,t)$ and $(x',t')$. Depending on their relation to the segment $\mc{S}=[(x,t),(x',t')]$
we divide soliton-antisoliton pairs into six disjoint classes (representative examples  are shown in Fig. \ref{fig:LRDI}):
The first class contains avoiding pairs with both rays lying on the right (RR) or on the left (LL) of segment $\mc{S}$
(an LL pair is shown in Fig. \ref{fig:LL}). These pairs do not affect the correlation.
We call left crossing (L) the pairs whose right going ray has a trajectory that crosses from the left and from below, while the left going ray of the same pair avoids the segment (see Fig. \ref{fig:L}). These pairs necessarily have velocities $v<v_s.$
We define in an analogous way right crossing pairs (R) which can have arbitrary velocity (see Fig. \ref{fig:R}).
The class of  double crossing pairs (D) includes pairs with both rays crossing $\mc{S}$  (see Fig. \ref{fig:D}). 
We call pairs such that the segment lies in between the two rays inclusions (I). 
Finally, a crucial role is played by those pairs whose right going ray crosses the segment ``from above'' (A) (see Fig. \ref{fig:Ltilde}). This is only possible if the velocity of the pair is greater than $v_s.$

\captionsetup[subfloat]{justification=justified,singlelinecheck=true,width=4cm}

\captionsetup[figure]{justification=justified,singlelinecheck=false}

The probabilities that a given pair belongs to either of these classes are calculated based on the uniform spatial distribution of the pairs, in the same way as in the previous sections. For example, the probability that a given pair belongs to class `L' for $t'>t$ is
\be
q_L =\frac1L \int_0^{\tilde v} \ud v 2vtf(v) + 
\frac1L \int_{\tilde v}^{v_s} \ud v [x'-x-v(t'-t)]f(v)
\ee
with $f(v)=f_{+-}(v)+f_{-+}(v)=2f_{+-}(v).$
The probabilities for the other classes are listed in the Appendix.

In a given configuration, the number of domains between  points $(x,t)$ and $(x',t')$ 
can be expressed in terms of  the number of pairs in each class. Counting the domains from the left, the point $(x,t)$ is in  domain 
\be
l=2n_{LL}+n_L+n_I+2n_A
\ee
while  $(x',t')$ is located in  domain $l'=2n_{LL}+2n_L+n_R+2n_D+n_I+n_A$. 
Thus the number of kinks between the two points is 
\be
s=l'-l=n_L+n_R+2n_D-n_A\,.
\label{s_general}
\ee
Let us now focus on the average over the charges of these $s$ kinks. We first remark that in this universal semiclassical limit 
kinks are strongly correlated:  the charges of kinks $2k-1$ and $2k$ always add up to 0. Therefore, 
if $s$ is odd we have to average over the charge of a single kink without a pair, yielding  $(e^{i2\pi\alpha/\beta}+e^{-i2\pi\alpha/\beta})/2=\cos(2\pi\alpha/\beta).$ If, on the other hand,  $s$ is even, then there are two possibilities: either both $l$ and $l'$ are even, so that all $s$ kinks form $s/2$ pairs and $\Phi_{l'}-\Phi_l=0,$ or both $l$ and $l'$ are odd, in which case we have $s/2-1$ pairs with zero total charge, and two unpaired kinks. Averaging over the charges of these two kinks yields $\cos^2(2\pi\alpha/\beta)\,.$ We have to separate, however, the special case, $s=0$:  then $\Phi_{l'}-\Phi_l=0$ irrespectively of the parity of $l.$ Notice that the $s=0$ configurations can be non-trivial due to the existence of the special cuts from above (A) that shift back the domain of 
$\Phi(x',t')$ (see the sign in Eq.~\eqref{s_general} and Fig. \ref{fig:Ltilde}).

Since only the parity of $s$ and $l$ matters, averaging over the velocities and positions of the pairs translates into a multiple sum over the numbers of the various types of pairs weighted by the probability of such a configuration:
\begin{widetext}
\begin{multline}
\tilde C_\alpha(x'-x;t,t') = \sum_{n_R,n_L,n_A,n_D,n_I}
\frac1{n_R!n_L!n_A!n_D!n_I!}Q_R^{n_R}Q_L^{n_L}Q_A^{n_A}Q_D^{n_D}Q_I^{n_I}
e^{-Q_R-Q_L-Q_A-Q_D-Q_I}\\
\Bigg(\delta_{s,0}\cdot1+(1-\delta_{s,0})\cdot
\left\{ 
\frac{1-(-1)^s}2 \cos\left(\frac{2\pi\alpha}\beta\right)+
\frac{1+(-1)^s}2\left[\frac{1+(-1)^l}2\cdot1+
\frac{1-(-1)^l}2\cos^2\left(\frac{2\pi\alpha}\beta\right)\right]
\right\}\Bigg)\,,
\end{multline}
where 
\be
\tilde C_\alpha(x'-x;t,t') \equiv  C_\alpha(x'-x;t,t')/ \mc{C}^{(0)}_\alpha(x'-x;t'-t) 
\ee
is the correlation function normalized by the vacuum correlation, and $Q_\mu(\Delta x;t,t')=Nq_\mu(\Delta x;t,t')$ with the probabilities $q_\mu$ listed in the Appendix.

The terms not multiplied by the Kronecker $\delta_{s,0} $ can be written, collecting the sign factors and using some basic trigonometric identities, as
\begin{multline}
\tilde C_1
=\cos^4\left(\pi\alpha/\beta\right) + \sin^4\left(\pi\alpha/\beta\right)e^{-2Q_R-2Q_L-2Q_A}+
\sin^2\left(\pi\alpha/\beta\right)\cos^2\left(\pi\alpha/\beta\right)
e^{-2Q_I}(e^{-2Q_L}+e^{-2Q_R-2Q_A})\,.
\end{multline}
The terms proportional to $\delta_{s,0}$ can be dealt with by using the integral representation for the Kronecker delta,
$
\delta_{s,0}=\int_{-\pi}^\pi\frac{\ud \phi}{2\pi} e^{i s\phi}$, and yield a contribution
\begin{multline}
\tilde C_2 = 2\sin^2\left(\pi\alpha/\beta\right)\cos^2\left(\pi\alpha/\beta\right)
\int_{-\pi}^\pi\frac{\ud \phi}{2\pi}e^{(e^{i\phi}-1)Q_R+(e^{2i\phi}-1)Q_D+(e^{-i\phi}-1)Q_A}
\left(e^{(e^{i\phi}-1)Q_L}-e^{-(e^{i\phi}+1)Q_L-2Q_I}\right)\,.
\end{multline}
The total expression for the correlator normalized by its vacuum value is then expressed as
\begin{multline}
\label{res}
\tilde C_\alpha(x'-x;t,t')=\tilde C_1+\tilde C_2\\
=\cos^4\left(\pi\alpha/\beta\right) + \sin^4\left(\pi\alpha/\beta\right)e^{-2Q_R-2Q_L-2Q_A}+
\sin^2\left(\pi\alpha/\beta\right)\cos^2\left(\pi\alpha/\beta\right)
e^{-2Q_I}(e^{-2Q_L}+e^{-2Q_R-2Q_A})\\
+2\sin^2\left(\pi\alpha/\beta\right)\cos^2\left(\pi\alpha/\beta\right)
e^{-(Q_L+Q_R+Q_D+Q_A)}
\int_{-\pi}^\pi\frac{\ud \phi}{2\pi}e^{Q_Re^{i\phi}+Q_De^{2i\phi}+Q_Ae^{-i\phi}}
\left(e^{Q_Le^{i\phi}}-e^{-Q_Le^{i\phi}-2Q_I}\right)\,.
\end{multline}
We remark that expanding  to second order in $\alpha$ and differentiating with respect to $x$ and $x'$ yields the topological charge density correlation function.

\end{widetext}

\section{Relaxation of correlation functions: Discussion of the result}

We shall now analyze the general result in Eq. \erf{res} and examine its behavior in various limits. 
First we note that setting $\alpha=\beta/2+k\pi$ only the second term of Eq. \erf{res} survives, and Eq. \erf{res}
simplifies to
\be
\tilde C_{\beta/2}(x'-x;t,t') = e^{-2Q_R-2Q_L-2Q_A}\,,
\ee
and we recover the result \erf{beta2res} with $Q=Q_R+Q_L+Q_A$, defined earlier in Eq. \erf{Qdef}.
For  $\alpha=\beta+k\pi$,  on the other hand, only the first term remains, and  the semiclassical calculation yields
 $\tilde C_{\beta}(x,x';t,t') = 1$.

Let us  now analyze the local correlation function, the equal time correlation function, and the asymptotic dynamic two-point functions for late times for general values of $\alpha$. 

\subsection{Local correlation function}

The local correlation function is obtained by setting $x\to x'.$ Using the special limits  of the $Q$ 
 listed in Eq. \erf{Qauto} of the Appendix, we obtain
\begin{widetext}
\begin{multline}
\label{resauto}
\tilde C_\alpha(0;t,t')= 
\cos^4\left(\pi\alpha/\beta\right) + \sin^4\left(\pi\alpha/\beta\right)e^{-\Delta t/\tau}\\
+\sin^2\left(\pi\alpha/\beta\right)\cos^2\left(\pi\alpha/\beta\right)
\left[e^{-t/\tau}(1+e^{-\Delta t/\tau})+2e^{-\Delta t/(2\tau)}(1-e^{-t/\tau})I_0\left(\frac{\Delta t}{2\tau}\right)\right]\,,
\end{multline}
\end{widetext}
where $\Delta t=t'-t$ and $\tau$ is the characteristic relaxation time defined in Eq. \erf{taudef}.

\begin{figure}[!t]
  \centering
   \includegraphics[width=.45\textwidth]{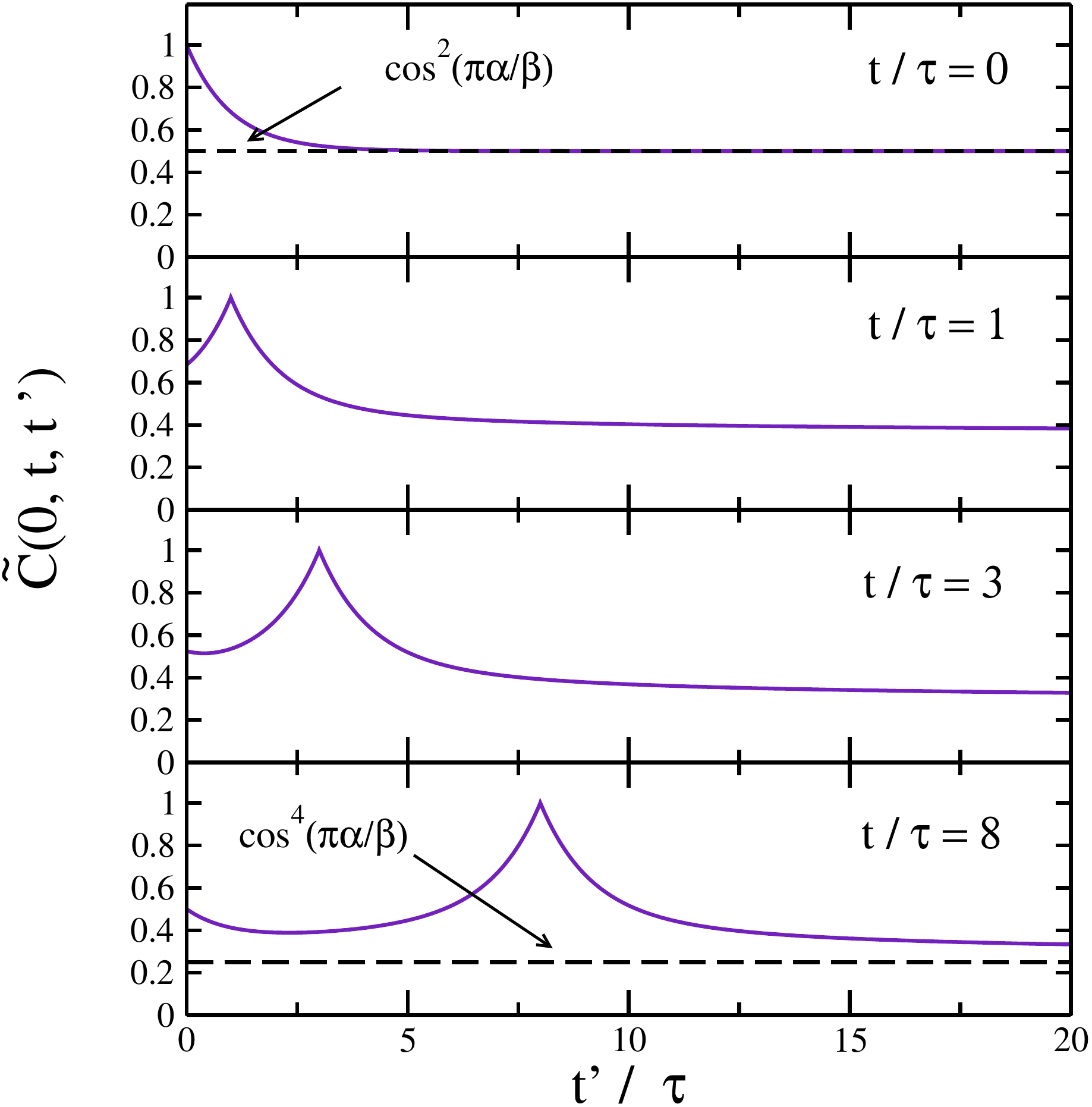}\label{fig:autocorr}
  \caption{Local correlation function for different values of $t$ as a function of $t'/\tau$, where $\tau$ is defined in Eq. \erf{taudef}. The decay is exponential for $t=0$ and diffusive $\sim1/\sqrt{|t'-t|}$ for all $t>0$.}
\label{fig:auto}
\end{figure}

The local correlation function is computed for a velocity distribution 
\be
\label{fv}
f(v) = \frac{4 v^2 }{v_0^3 \sqrt{\pi}}  \,e^{-v^2/v_0^2}
\ee
and displayed  in Fig. \ref{fig:auto}. This form of $f(v)$ is motivated by the observation that the amplitude 
$K(\theta)$ is an odd function of $\theta$ around $\theta=0\;$ \footnote{This is due to the relation $K_{m_2',m_1'}(\theta)=S^{m_2',m_1'}_{m_1,m_2}(2\theta)K_{m_1,m_2}(-\theta)$ which follows from the exchange relation of the creation operators in Eq. \erf{psi0}. Here $S(\theta)$ is the S-matrix having the universal low energy limit \erf{lowS} which implies $K_{m_1,m_2}(\theta)=-K_{m_2,m_1}(-\theta)$ as $\theta\to0.$}.

It is easy to check that $\tilde C_\alpha(0;t,t)=1$ as $\Delta t\to 0$, 
a condition giving rise to the spike structures in Fig. \ref{fig:auto}.
 The most interesting feature of the result is its behavior for large time separation. For a finite and fixed $t$
 and large values of $t'$  we find 
\begin{multline}
\tilde C_\alpha(0;t,t'\to\infty)
=\cos^4\left(\pi\alpha/\beta\right) \\
+\sin^2\left(\pi\alpha/\beta\right)\cos^2\left(\pi\alpha/\beta\right)
\Bigl[e^{-t/\tau}+\frac{2(1-e^{-t/\tau})}{\sqrt{\pi\Delta t/\tau}}\Bigr]\,.
\end{multline}
 For any $t>0,$ the late time behavior is thus {\em diffusive}, similar to the behavior found within the semiclassical approximation in thermal equilibrium \cite{Damle2005,Rapp2006,Rapp2008}.
As $\Delta t\to\infty,$ the local correlation function approaches a $t$-dependent non-zero constant, equal to the product of 
expectation values given in Eq. \erf{1ptres} at times $t$ and $t'\to\infty,$ as expected. (The normalization factor given by the vacuum correlator also factorizes in this limit.)

The origin of the  diffusive contribution  can be understood  in terms of the simple picture of randomly diffusing domains, 
discussed in Sec.~\ref{sec:exp}
as follows: for sufficiently long times $t\gg \tau$ the point $(0,t)$ is in a randomly selected domain $l$. 
This domain follows a diffusive motion, and remains at the point  $x=0$ with probability $\sim 1/\sqrt{D\Delta t}$, with 
$D\sim 1/(\tau \rho^2)$ the diffusion constant. With this probability, the phase of the vertex functions 
at $(0,t)$ and $(0,t')$ is exactly the same, and the correlator gives 1.

Interestingly, the diffusive term vanishes only for $t=0$ in which case 
we recover the exponential behavior found for the expectation value:
\be
\tilde C_\alpha(0;0,t')
=\cos^2\left(\pi\alpha/\beta\right) + \sin^2\left(\pi\alpha/\beta\right)e^{-t'/\tau}\,.
\ee
It is also interesting to note that for $\alpha=\beta/2+k\pi,$
\be
\tilde C_{\beta/2}(0;t,t') =e^{-\Delta t/\tau}
\ee
is independent of the time after the quench. This is true also for more general initial states having pairs with non-zero total charge, as can be seen from Eqs. (\ref{Qdef},\ref{beta2res}) where the first two terms in Eq. \erf{Qdef} vanishes upon $v_s,\tilde v\to0.$ This instant relaxation might be an artefact of the semiclassical approximation. However, the calculation for $\alpha=\beta/2$ is equivalent to that in the Ising field theory which can be mapped to free fermions, and in such systems similar behavior has already been observed. In Ref. \onlinecite{Kormos2014},  for a non-relativistic gas of bosons  the exact density-density correlation function was obtained analytically after quenching the contact interaction strength from zero to infinity, exploiting the mapping between the infinitely repulsive Bose gas (Tonks--Girardeau gas) and free fermions. The local correlation function was shown exactly to be  time-independent. Similar behavior was found for relativistic free field theories in a different situation, after connecting two semi-infinite systems thermalized at different temperatures \cite{Doyon2014b}. Thus in the case of the Ising model, the time independence of the spin-spin local correlation function found in the semiclassical approach can turn out to be an exact result.

\subsection{Equal time correlation function}

Let us turn now to the relaxation of the equal time correlation function. The $Q$ functions for $t=t'$ are given in Eqs. \erf{Qequal}. Since $Q_A=0,$ the $\phi$-integral in Eq. \erf{res} can easily be evaluated and the correlation function becomes

\begin{multline}
\label{resequal}
\tilde C_\alpha(\Delta x;t,t)
 = \cos^4\left(\pi\alpha/\beta\right) + \sin^4\left(\pi\alpha/\beta\right)e^{-4\rho\Delta x+4Q_D}+\\
 2\sin^2\left(\pi\alpha/\beta\right)\cos^2\left(\pi\alpha/\beta\right)
\left[e^{Q_D-2\rho\Delta x}+e^{-t/\tau}(1-e^{-Q_D}])\right]\,,
\end{multline}
where 
\be
Q_D(\Delta x,t)=\rho\int_0^{\tilde v}\ud v(\Delta x-2vt)f(v)\,.
\ee
It can easily be shown that for $\Delta x=0$ ($Q_D=0$) we recover again $\tilde C_\alpha(0;t,t)= 1.$

\begin{figure}[!t]
  \centering
   {\includegraphics[width=.45\textwidth]{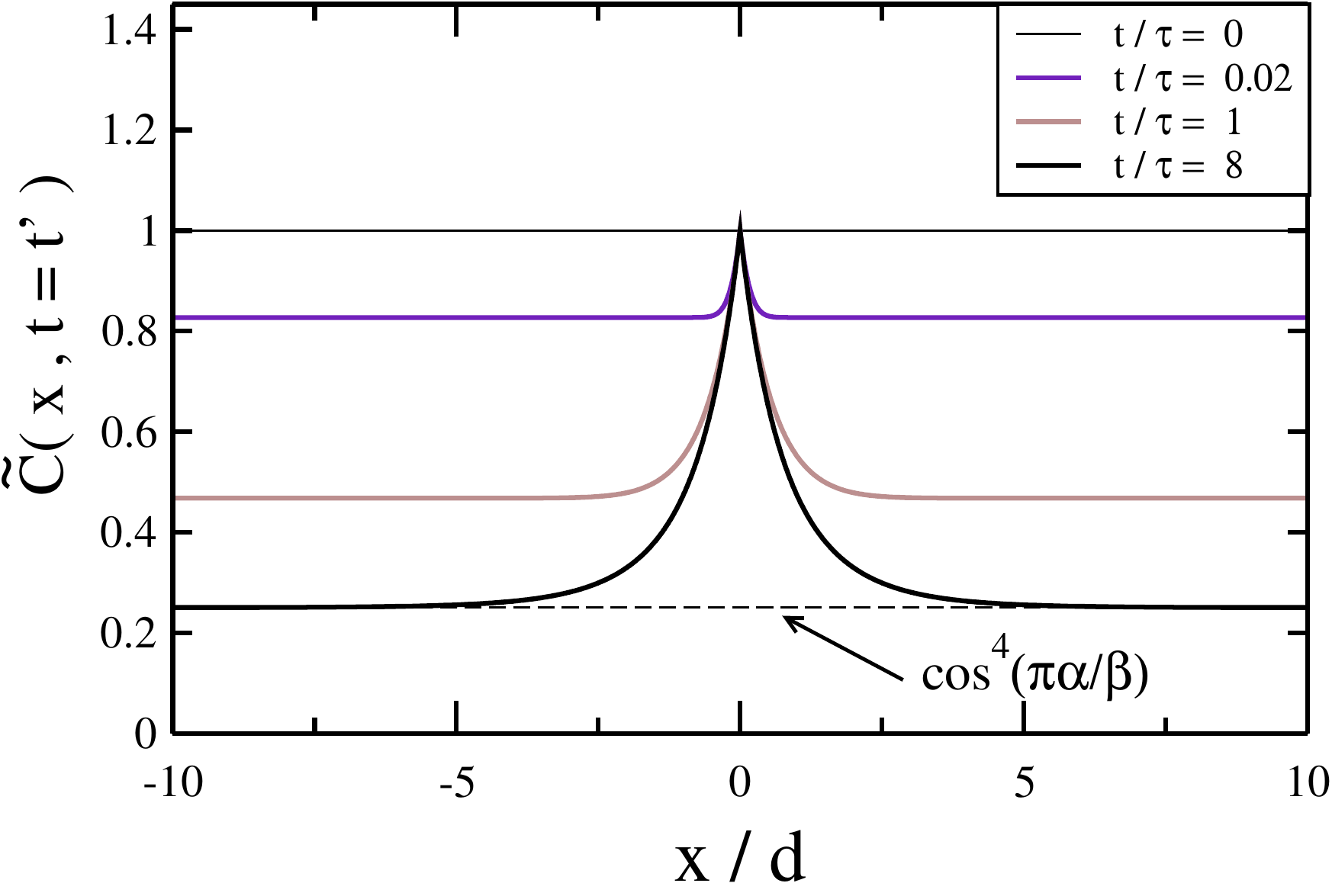}}
   \caption{Equal time correlation funtion at different times as a function of $x/d$ where $d=1/(2\rho)$ is the mean interparticle spacing.}
\label{fig:equal}
\end{figure}

The correlation function is plotted in Fig. \ref{fig:equal} using the distribution in Eq. \erf{fv}. At any finite $t>0$, as $\Delta x\to\infty$ it  approaches exponentially
a constant,
\be
\tilde C_\alpha(\Delta x\to\infty;t,t)
=\left[\cos^2\left(\pi\alpha/\beta\right)+\sin^2\left(\pi\alpha/\beta\right)e^{-t/\tau}\right]^2\,.
\ee
This is the square of the expectation value \eqref{1ptres} at time  $t$, so the cluster decomposition property holds for all times after the quench. The connected part of the correlation function is thus exponential $\sim e^{-\rho\Delta x}$ with correlation length given by the density of kink pairs.

A different result is obtained by taking  the $t\to\infty$ limit first, where we obtain  the asymptotic steady state correlation function,%
\be
\tilde C_\alpha(\Delta x;t\to\infty) 
= \left[\cos^2\left(\pi\alpha/\beta\right)+\sin^2\left(\pi\alpha/\beta\right)e^{-2\rho\Delta x}\right]^2\,.
\ee
The correlation length in the asymptotic state is thus $\xi_\text{as}=1/(2\rho)$ rather than $1/\rho,$ showing that 
the large separation and large time limits do not commute.

\subsection{Asymptotic steady state}
Finally, let us discuss the correlation function in the asymptotic steady state.
Asymptotically, time translational invariance is restored, and 
the probabilities given in Eqs. \erf{Qas} become functions of the spatial and temporal separations only. 
Consequently, the steady state correlation function can be expressed as
\begin{widetext}
\begin{multline}
\tilde C^{as}_\alpha(\Delta x;\Delta t)\equiv\lim_{t,t'\to\infty}\tilde C_\alpha(\Delta x;t,t')\\
=\cos^4\left(\pi\alpha/\beta\right)
+\sin^4\left(\pi\alpha/\beta\right)e^{-4\rho\Delta x-4Q_A}
+2\sin^2\left(\pi\alpha/\beta\right)\cos^2\left(\pi\alpha/\beta\right)e^{-2\rho\Delta x-2Q_A}
I_0\left(2\sqrt{Q^{\rm as}_A(Q^{\rm as}_A+2\rho\Delta x)}\right)\,.
\end{multline}
\end{widetext}
with
\be
Q^{\rm as}_A(\Delta x, \Delta t)=\rho\int_ {v_s}^\infty\ud v(v\Delta t-\Delta x)f(v)\,.
\ee
Assuming a power law behavior $f(v)=f_0 v^k+\mc{O}(v^{k+1})$ for small velocities, the asymptotic value is approached as
$\sim 1/t^{(k+1)}.$

For infinite separation, $\Delta x\to\infty,$  the cluster decomposition holds and we recover the square of the asymptotic value of the 1-point function,  
\be
C^{\rm as}_\alpha(\Delta x\to\infty;\Delta t) \to \mc{G}_\alpha^2\cos^4\left(\pi\alpha/\beta\right)=\vev{e^{i\alpha\Phi(x,\infty)}}^2\,.
\ee

For large temporal separation, $\Delta t/(4\tau)\gg\rho\Delta x,$ $Q_A\to\Delta t/(4\tau)-\rho\Delta x$, and to 
 leading order
\begin{multline}
\tilde C^{as}_\alpha(\Delta x;\Delta t\to\infty) = \cos^4\left(\pi\alpha/\beta\right)\\
+2\sin^2\left(\pi\alpha/\beta\right)\cos^2\left(\pi\alpha/\beta\right)\frac1{\sqrt{\pi\Delta t/\tau}}\,,
\end{multline}
that is, the diffusive behavior of the local correlation function is recovered.

\section{Summary and outlook}

In this paper we developed a semiclassical approach to  study the 
 time evolution of correlation functions  after small quenches in the repulsive regime of the sine--Gordon field theory. 
 Assuming that the velocity of quasiparticles (kinks) is very small compared to the gap and using the asymptotic S-matrix, we 
 were able to derive analytical approximations for the 1-point and general dynamical 2-point functions. 
 Remarkably, this simple method --- not restricted to integrable systems --- allowed us to recover results for the 1-point functions obtained earlier by means of  exact form factor expansions. Though these calculations were performed in the repulsive regime, 
we expect our results to remain  valid even in the attractive regime $\beta<1/\sqrt{2}$ where  breathers (bound states of solitons and antisolitons) are also present. Passing a breather does not change the value of $\Phi,$ and in the low energy limit the scattering between the breathers and the kinks is purely transmissive.  Thus  breathers decouple from  
kinks in the semiclassical limit and should not influence the correlation functions dramatically.

Our semiclassical calculations yield certain results that are somewhat surprising at first sight.
We find, e.g., that the expectation value  $\bigl<{ e^{i\alpha\Phi(x,t)}}\bigr>$ 
    approaches exponentially a constant value for generic $\alpha$'s (see Eq. \erf{1ptres}). 
This is a consequence of the fact that the low-energy limit of the S-matrix is perfectly reflective, implying 
  that the spatial sequence of the domains is conserved in time. Allowing also transmission in course of soliton-antisoliton scattering would break this pattern and lead to random domain sequences and a decay of the expectation value. 
 
We also obtained an analytic expression for the time evolution of the dynamic correlations $\bigl< e^{i\alpha\Phi(x,t)}e^{-i\alpha\Phi(x',t')}\bigr> $
of generic vertex operators and analyzed it in the various limits. We found that the cluster decomposition property holds during the non-equilibrium time evolution after the quench, at least for not very short times. The asymptotic value of the correlation functions is approached as a power law. The two-time correlations and the local correlation functions, in particular, show diffusive behavior. This can be understood heuristically based upon the picture  of `magnetic'  domains performing a random walk. 

Both the diffusive correlations and the saturating expectation values are  consequences of the assumption of  a perfectly reflective  S-matrix.  At any finite energy, however, the transmission has a small but finite probability  $P_\text{tr}\sim v^2/c^2$, 
yielding a finite time scale, above which expectation values should decay to zero and the diffusive behavior is also expected to turn into an exponential decay.  In this sense our asymptotic results describe a sort of prerelaxation plateau which eventually decays at late times. We can estimate relatively simply the corresponding time scale  by considerations similar 
to those in Ref.~\onlinecite{Rapp2006}. Within a time period $T$ a given kink participates in $T/t_\text{coll}$ collisions with a collision time  $t_\text{col}\propto\tau$. So after time $T\sim\tau c^2 /\bar v^2$ domains should change their color with a probability close to one and the reflective approximation of the S-matrix should break down. 
These considerations point into a possible direction of improving the present method by incorporating the leading non-reflective part of  the S-matrix, or to use the full S-matrix. These explorations need, however, extensive numerical simulations and are beyond the scope of the present work.

\acknowledgments
We acknowledge useful discussions  G\'abor Tak\'acs. We are especially grateful to 
 Pascu Moca, who verified through numerical simulations the predictions of our semiclassical 
 computations~\cite{Mocaetal}.  This research has been supported by the Hungarian
Scientific Research Fund OTKA under Grant No.
K105149 and by the EU Marie Curie IIF Grant PIIF-GA-2012- 330076.

\setcounter{equation}{0}
\renewcommand{\theequation}{A\arabic{equation}}

\onecolumngrid

\bigskip

\bigskip
\bigskip
\bigskip

\section*{Appendix}

In this Appendix we list the integral expressions for the probabilities that a given pair belongs to one of the classes defined in Sec. \ref{sec:general}. These are calculated similarly to the probabilities computed in Sec. \ref{sec:exp} and \ref{sec:corrbeta2}, based on the uniform spatial distribution of pairs giving rise to rays that intersect or avoid the segment connecting the two operator insertion points in space and time. We work with the convention $x'\ge x,t'>t$ and we use the notation 
\be
\tilde v = \frac{x'-x}{t'+t}\,,\qquad v_s = \frac{x'-x}{t'-t}\,.
\ee

\begin{itemize}
\item Probability that a pair leads to a double intersection
\be
q_D = \frac1L\int_0^{\tilde v}\ud v [x'-x-v(t'+t)]f(v)\,.
\ee
\item Probability that a pair leads to an inclusion
\be
q_I = \frac1L\int_{\tilde v}^{v_s}\ud v [v(t'+t)-(x'-x)]f(v)+
\frac1L\int_{v_s}^\infty\ud v 2vt f(v)\,.
\ee
\item Probability that a pair leads to one right intersection
\be
q_R =\frac1L \int_0^{\tilde v} \ud v 2vt'f(v) + 
\frac1L \int_{\tilde v}^{v_s} \ud v [x'-x+v(t'-t)]f(v)+\Theta(t'-t)\frac1L \int_{v_s}^\infty \ud v [x'-x+v(t'-t)]f(v)\,.
\ee
\item Probability that a pair leads to one left intersection
\be
q_L =\frac1L \int_0^{\tilde v} \ud v 2vtf(v) + 
\frac1L \int_{\tilde v}^{v_s} \ud v [x'-x-v(t'-t)]f(v)+\Theta(t-t')\frac1L \int_{v_s}^\infty \ud v [x'-x-v(t'-t)]f(v)\,.
\ee
\item Probability that a pair leads to one left intersection from above
\be
q_A = \Theta(t'-t)\frac1L \int_{v_s}^\infty \ud v [x-x'+v(t'-t)]f(v)\,.
\ee
\end{itemize}
In the correlation functions these probabilities appear multiplied by the total number of pairs $N$, e.g.
\be
Q_D = Nq_D=\rho\int_0^{\tilde v}\ud v [x'-x-v(t'+t)]f(v)\,.
\ee

Below we list some limiting forms of these expressions.

\begin{enumerate}

\item For the local correlation function $x=x',$ so both $\tilde v=v_s=0,$ and
\be
\label{Qauto}
Q_D=Q_L=0\,,\qquad Q_R=Q_A=\frac{t'-t}{4\tau}\,,\qquad Q_I=\frac{t}{2\tau}\,,
\ee
where
\be
\tau^{-1}\equiv 4\rho\int_0^\infty\ud v v f(v)\,.
\ee

\item For the equal time correlation function $t=t',$ the velocities become $\tilde v=\Delta x/(2t),$ $v_s=\infty,$ and
\begin{subequations}
\label{Qequal}
\begin{alignat}{2}
Q_D&=\rho\int_0^{\tilde v}\ud v(\Delta x-2vt)f(v)
\,,\qquad&Q_I&=\rho\int_{\tilde v}^\infty\ud v(2vt-\Delta x)f(v)=\frac{t}{2\tau}-\rho\Delta x
+Q_D\,,\\
Q_A&=0\,,\qquad&Q_L&=Q_R=2t\,\rho \int_0^{\tilde v}\ud v vf(v)+\Delta x\,\rho\int_{\tilde v}^\infty\ud vf(v)=\rho\Delta x
-Q_D\,,
\end{alignat}
\end{subequations}
where we used $\int_0^\infty\ud v f(v)=1.$

\item The asymptotic steady state is obtained by taking $t,t'\to\infty$ with $\Delta t$ fixed, when we have $\tilde v\to0,$ and (for $t'\ge t$)
\begin{subequations}
\label{Qas}
\begin{alignat}{3}
Q_A&=\rho\int_ {v_s}^\infty\ud v(v\Delta t-\Delta x)f(v)\,,\qquad&Q_D&\to0\,,\qquad& Q_I&\to\infty\,,\\
Q_R&=\rho\Delta x+\Delta t/(4\tau)\,,\qquad& Q_L&=\rho\Delta x-\Delta t/(4\tau)+Q_A\,.&&
\end{alignat}
\end{subequations}
The leading correction to these asymptotic expressions for large $t$ comes from integrals of the type $\int_0^{\tilde v}\ud v f(v)(a+b v t).$ Assuming a power law behavior $f(v)=f_0 v^k+\mc{O}(v^{k+1})$ for small velocities, this gives a $\sim 1/t^{(k+1)}$ power law correction.

\end{enumerate}

\twocolumngrid

\bibliographystyle{apsrev4-1}

\bibliography{sGpaper_arxiv}

\end{document}